# Electrical Conductivity of Copper-Graphene (Cu-Gr) Composites: The Underlying Mechanisms of Ultrahigh Conductivity

Jiali Yao[1], Uschuas Dipta Das[2], Hamid Safari[2], Md Ashiqur Rahman Laskar[3], Junghoon Yeom[4], Umberto Celano[3], and Wonmo Kang[1, 2,*]

[1]Materials Science and Engineering, Fulton Schools of Engineering, Arizona State University, Tempe, AZ 85287, USA
[2]Mechanical Engineering, Fulton Schools of Engineering, Arizona State University, Tempe, AZ, USA
[3]School of Electrical, Computer and Energy Engineering, Arizona State University, Tempe, AZ, USA
[4]Multifunctional Materials Branch (Code 6350), U.S. Naval Research Laboratory, Washington, DC, USA

*wonmo.kang@asu.edu

**Abstract**
Copper-graphene composite (CGC) conductors are widely considered as a potential alternative to pure copper (Cu). Yet, the effect of graphene (Gr) on the electrical conductivity of CGCs remains elusive, and their electrical performance is still controversial. This work addresses these unresolved questions by unambiguously quantifying how the electrical properties of CGCs depend on the characteristics of Gr and Cu. Gr is synthesized on Cu foils, foams, and wires by utilizing a wide range of chemical vapor deposition conditions to independently control their characteristics. Then the Gr-enhanced electrical conductivity ($\Delta\sigma$) is characterized for CGCs with different Cu geometries and Gr qualities. This study confirms that unprecedented electrical conductivity ($\Delta\sigma$ = 17.1%) can be achieved only when both Gr and Cu are carefully optimized. Specifically, the study reveals three key factors: (1) $\Delta\sigma$ is positively correlated with continuity of Gr; (2) CGCs with a continuous monolayer Gr exhibit a strong $\Delta\sigma - A_s$ linear relation where $A_s$ is the specific surface area of a CGC; and (3) $\Delta\sigma$ becomes more pronounced when a Cu matrix has a curved cross-section. This work reveals the fundamental mechanisms of how Gr influences the overall electrical conductivity of CGCs and, therefore, is a crucial step toward designing and manufacturing high-performance CGC conductors for emerging applications.

## 1. Introduction

The demand on electricity is higher than ever due to the emerging technologies including electric vehicles [1], cryptocurrency mining [2], and machine learning models training for artificial intelligence [3]. However, the transmission and distribution of electricity remains relatively inefficient – the global energy loss is around 16% [4] – mainly due to Joule heating from the electrical resistance of conductors. Therefore, improving electrical conductivity of conventional copper (Cu)-based conductors is of great importance. Graphene (Gr) [5], a two dimensional (2D) carbon allotrope with honeycomb lattice, has ultrahigh electron mobility ( $> 200\,000\ cm^2v^{-1}s^{-1}$ ) [6], thousands-times-higher than Cu ( $\approx 43\ cm^2v^{-1}s^{-1}$ ) [7]. In addition, the in-plane electrical conductivity of free-standing Gr achieves 100 $MS/m$ [8], significantly higher than both copper (the international annealed copper standard (IACS): 58.1 $MS/m$) [7] and silver (63.0 $MS/m$) [7]. Because of these outstanding electrical properties, Gr has been considered as a promising candidate to improve the electrical conductivity of Cu, arguably the most common conductor, by synthesizing Cu-Gr composites (CGCs) [9, 10].

Several techniques have been developed to synthesize CGC conductors, including chemical vapor deposition (CVD) [11, 12], powder metallurgy [13], molecular level mixing [14], etc. Among them, CVD-grown Gr on a Cu substrate, in which gaseous organic molecules are catalytically decomposed and graphitized on the surface of Cu (see **Figure S1** in the supporting information for the schematics) [15], offers three attractive advantages. Firstly, CVD is known for the synthesis of large-area, uniform, and high-quality Gr [16]. The defect density of CVD-grown Gr is comparable with the best reported exfoliated Gr [17] and much lower than those of reduced Gr oxide (rGO) and Gr oxide (GO) [9]. Secondly, the gas-phase reaction enables sufficient matter transport and continuous conformal coverage of Gr on large Cu surfaces regardless of matrix geometry [16]. Lastly, a-few-layer Gr grown by CVD can deviate from the AB-stacking sequence (also known as Bernal stacking) that is found in bulk phase [18, 19]. As the interlayer coupling under AB-stacking was found to hinder in-plane electron transport, the CVD-grown "decoupled" a-few-layer Gr may have higher electrical conductivity than the AB-stacked counterpart [20]. Owing to these unique advantages, significant technical advances in CGCs (i.e., electrical conductivity > 100% IACS) have been achieved by CVD-based CGCs. For example, Cao et al. [11] reported an electrical conductivity of 117% IACS with only a 0.008 $vol.\%$ of Gr in a Cu-foil-based CGC. Furthermore, Kashani et al. [12] developed an axially bi-continuous Gr-coated Cu wire with 123% IACS.

Despite the recent progress in CGC conductors, the effect of graphene (Gr) on the electrical conductivity of CGCs remains elusive, and their electrical performance is still controversial. For example, CGCs with a broad range of the electrical conductivity, ranging from 10% to 123% IACS, has been reported in the literature [9]. Moreover, two opposite views on the effect of Gr on CGC conductors are available. A first view is that Gr enhances electrical conductivity of CGC, e.g., an enhanced electrical and thermal conduction in Gr encapsulated nanowires was observed and attributed to inelastic electron scattering at the Gr-Cu interface [21]; an ultrahigh electrical conductivity of Gr embedded Cu foils were explained by the electron doping of Gr by Cu [11] where electron transfer from Cu to the unoccupied states of Gr was predicted by the first-principles calculations. In contrast, a recent work [22] based on the first-principles method reported that electron scattering increases at the Gr-Cu interface, despite the increased carrier concentration, and, therefore, the conductivity decreases as the carbon ratio increases.

This work unambiguously resolves the current controversy by quantifying Gr-enhanced electrical conductivity of CGCs while precisely controlling the important characteristics of both Gr and Cu, such as their volume fraction, geometry, and spatial distribution. In this regard, we perform parametric studies of CVD Gr growth using a different gas flow rate and growth time on three different types of Cu substrates (i.e., foil, wire, and foam). Then the electrical conductivity of each CGC is measured by 4-point probe measurements. This study clearly indicates that the electrical conductivity of CGC is determined by the characteristics of both Gr and Cu, including the quality and continuity of graphene, a Gr-to-Cu volume fraction, and geometry of Cu. In other words, the conductivity of CGC can be either higher or lower than that of pure Cu depending on the specific characteristics of both Gr and Cu.

One of the key findings is the importance of graphene morphology in CGC – continuous monolayer graphene in CGC results in the best electrical performance. Using optimized CVD parameters, 1.04%, 14.1%, and 17.1% improvement in electrical conductivity have been observed for foil-, foam-, and wire-based CGCs, respectively. To explain the quantitatively different enhancements, we carefully consider the geometry of Cu matrices and reveal two important correlations between the overall electrical conductivity of CGC and the geometry of Cu. The main conclusion is twofold: first, the electrical conductivity improvement is linearly proportional to the specific surface area of Cu matrix; second, curved surface (wires or foams) benefits more from Gr coating than flat surface (foil). Apart from the scientific findings, our work is also relevant to practical applications of CGCs in two possible ways: (1) CGCs with specifically tuned electrical conductivity may be possible by tailoring the characteristics of Gr and Cu and (2) high-throughput manufacturing of high-performance CGC conductors may be achieved by using macroscopic Cu-foam based CGC (e.g., via roll-to-roll process).

## 2. Results and Discussion
### 2.1 Characterization of CVD grown graphene on Cu

CVD Gr growth was performed on three different types of Cu substrates, i.e., foil, wires and foams, to investigate Gr growth on a flat surface, cylindrical surface, and three-dimensional (3D) network. **Figure 1** shows the three types of CGCs, namely foil-, wire-, and foam-based CGCs. Note that continuous Gr coating is not visible under ordinary optical microscopic (OM) and scanning electron microscopic (SEM) imaging conditions on all the Cu substrates due to its atomically thin nature. For a foil-based CGC in **Figure 1**a, the large grains, ranging from tens to hundreds of micrometers, can be clearly observed due to thermal grain growth during CVD at about 1000 °C. Interestingly, parallel patterns, likely formed during the production of Cu foil (e.g., by rolling), are still visible after CVD (see the white arrows in **Figure 1**a; more details can be found in **Figure S2**a in supporting information). The inset of **Figure 1**a shows the cross section of a 28-$\mu m$-thick foil-based CGC. **Figure 1**b-d show the SEM images of wire-based CGCs with three different nominal diameters of 80, 25, and 10 $\mu m$. For all cases, the grain boundaries form dimples on the wire surfaces and make the diameter slightly smaller than its nominal diameter. The insets in **Figure 1**b-d are the cross-section of the corresponding wire, showing that the wires' circular geometry is well reserved after CVD. **Figure 1**e shows the overview of a foam-based CGC consisting of randomly oriented micro-ligaments because a commercially available Cu foam, produced by a sponge-replication method [23], was used as a substrate for graphene growth. The cross section of individual ligaments is hollow as shown in **Figure 1**f. An SEM image in **Figure 1**g shows details of a ligament in a foam-based CGC, e.g., uneven surface profiles (see the ripples indicated by white arrows in **Figure 1**g). Based on these images, the outer dimension of a ligament is about 50 $\mu m$ with a wall thickness of about 10 $\mu m$ (more details can be found in Section 2.3).

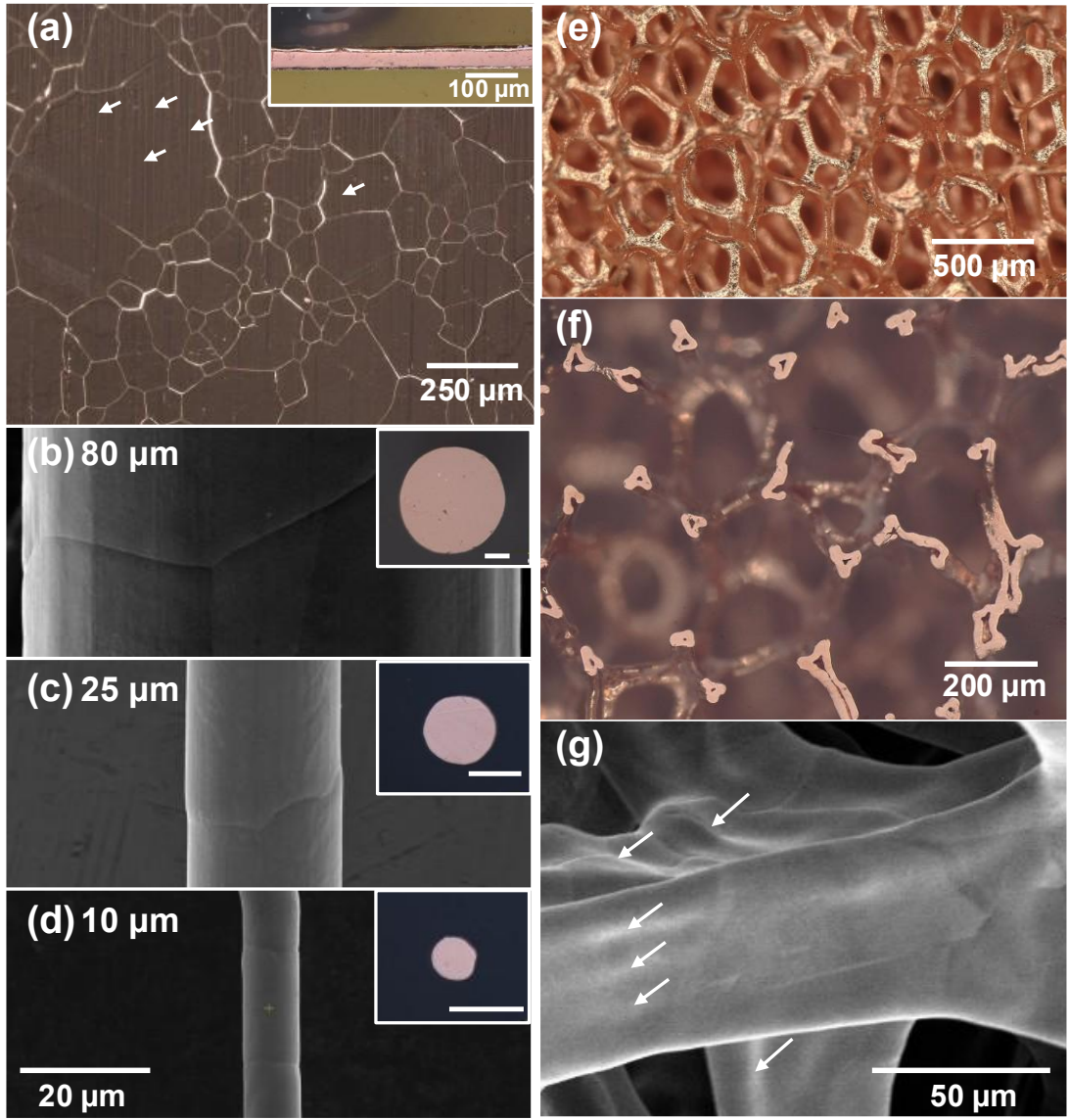

**Figure 1**. Foil-, wire- and foam-based copper graphene composites (CGCs) where Gr is invisible on the surface of Cu substrates under ordinary scanning electron spectroscopic (SEM) or optical microscopy (OM) imaging conditions. (a) An OM image of the top view of the Cu foil after CVD. The Cu grains have various sizes ranging from tens to hundreds of micrometers. Parallel lines are observed and some of the lines are indicated by white arrows. The inset is the cross-section of a 28-μm-thick Cu foil. (b)-(d) SEM images of wire-based CGCs with three different nominal diameters (80 μm, 25 μm and 10 μm). The insets in (b)-(d) are the polished cross-section of the CGC wires, showing their circular geometry after CVD. All scale bars in (b)-(d) are 20 $\mu m$. (e)-(g) The three-dimensional (3D) structure of the foam-based CGCs. Randomly oriented ligaments are observed in two OM images, (e) top view and (f) cross-section view. Note that each ligament has a hollow-tube structure. (g) An SEM image of ligaments, showing the surface topography of the Cu ligaments. In (g), the white arrows indicate the ripples of the surface.

  First, foil-based CGCs were selected for characterizing the CVD-process-dependent quality of graphene because of their simple planar geometry unlike wire- and foam-based CGCs. For quantitative measurements, CVD-grown graphene films were synthesized under the growth times ($t_{CVD}$) of 10 $s$, 1, 5, 10, and 20 $min$ at the same flow rate (i.e., 1 $sccm$) of benzene and then transferred onto a silicon wafer coated by a 300-nm-think silicon oxide layer (hereafter SiO$_2$/Si) for detailed optical observations. Further details on graphene transfer technique can be found in **Figure S3**, supporting information.

For clarity, the 1$^{st}$, 2$^{nd}$, and 3$^{rd}$ columns of **Figure 2** are corresponding to foil-based CGCs with $t_{CVD} = 10\ s$, $5\ min$, and $20\ min$, respectively. **Figure 2**a-c shows the optical images of Gr on SiO$_2$/Si from foil-based CGCs. For $t_{CVD} = 10\ s$, isolated microscale Gr islands can be observed where their lateral size is about 1~5 μ$m$ (see the insets of **Figure 2**a and **Figure S4**b, supporting information). A small dark spot can be seen near the center of each Gr island, known to be the nucleation center consisting of bi- or multilayer Gr [24]. These features can be observed in the upper right inset of **Figure 2**a, a low-voltage SEM image taken before Gr transfer (also see **Figure S4**b in supporting information). This result confirms that the transfer process does not distort Gr morphology. For $t_{CVD} = 5$ and $20\ min$, the light purple color (see the dashed white arrows in **Figure 2**a; see more details in **Figure S2**, supporting information) from the bare SiO$_2$/Si surface is no longer observable, suggesting that the entire surface is fully covered by Gr. However, the case of $t_{CVD} = 5\ min$ shows relatively uniform color distribution compared to both $t_{CVD} = 10\ s$ and $= 20\ min$. Note that dark blue areas in **Figure 2**c (see the white arrow) suggest multilayer Gr structures near preferred nucleation sites, e.g., along rolling patterns and Cu grain boundaries [25, 26] (see **Figure S2**, supporting information).

To confirm our optical observations, we also performed Raman mapping over 20 μ$m$ by 20 μ$m$ areas on the same Gr samples shown in **Figure 2**(a)-(c). In the Raman spectroscopy of Gr [27, 28], the intensity ratio of the $D$ peak ($I_D \approx 1350\ cm^{-1}$) to the $G$ peak ($I_G \approx 1580\ cm^{-1}$) as well as the $2D$ peak ($I_{2D} \approx 2700\ cm^{-1}$) to $I_G$ are highly indicative of Gr quality. For example, a larger $I_D/I_G$ ratio indicates a higher defect density (e.g., Gr edges), while $I_{2D}/I_G > 2$, $\approx 1$, and $< 1$ correspond to mono, bilayer, and multilayer Gr, respectively. Considering this, the Raman maps showing $I_{2D}/I_G$ in **Figure 2**d-f match well with the optical observations. For example, **Figure 2**d shows small blue areas, indicating multilayer Gr, surrounded by mono/bilayer Gr (green and light blue). One discrepancy is that the Raman map in **Figure 2**d does not clearly show the discrete characteristics of isolated Gr islands. This is due to the relatively large laser spot size ($\approx 2.3$ μ$m$, see **Figure S4**a in the supporting information). As shown in **Figure S4**b, the gaps between isolated Gr islands are typically smaller than the laser spot size and, therefore, the discrete characteristics cannot be accurately resolved. For $t_{CVD} = 5$ and $20\ min$, the similarities from optical and Raman analyses becomes obvious, including uniform color distribution (light blue in **Figure 2**b and green in **Figure 2**e) and the nearly one-to-one matching patterns between **Figure 2**c and f. Note that these results reveal that $t_{CVD} = 5\ min$ results in uniform monolayer graphene unlike the partial coverage of Gr for $t_{CVD} = 10\ s$ and inhomogeneous coverage by mono- to multilayer Gr for $t_{CVD} = 20\ min$.

To support our conclusions with statistical relevance, **Figure 2**g-i and **Figure 2**j-l show the number of pixels versus the $I_{2D}/I_G$ ratio and $I_G$, respectively, from the 2D Raman mapping. For $t_{CVD} = 10$ s, the counts distribute in the range of $I_{2D}/I_G = 1$~3.7, where the peak occurs at $I_{2D}/I_G \approx 1.8$, indicating that Gr consists of monolayer and bilayers, but the monolayer is dominant. For $t_{CVD} = 5\ min$, the peak shits to $I_{2D}/I_G \approx 3.1$, indicating the weight of monolayer Gr becomes more significant, compared to the Raman signals from the nucleation centers. This suggests that the expansion of dominant monolayer Gr prevails until the full surface coverage. For $t_{CVD} = 20\ min$, the $I_{2D}/I_G$ distribution becomes bimodal with peaks at $\approx 0.9$ and $\approx 2.8$ showing location-dependent Gr structures. The peak at $\approx 0.9$ can be attributed to the precipitation of additional nucleation sites beneath the dominant layer. **Figure S5** in the supporting information shows Raman spectra of 10 randomly selected spots in a larger area for additional details.

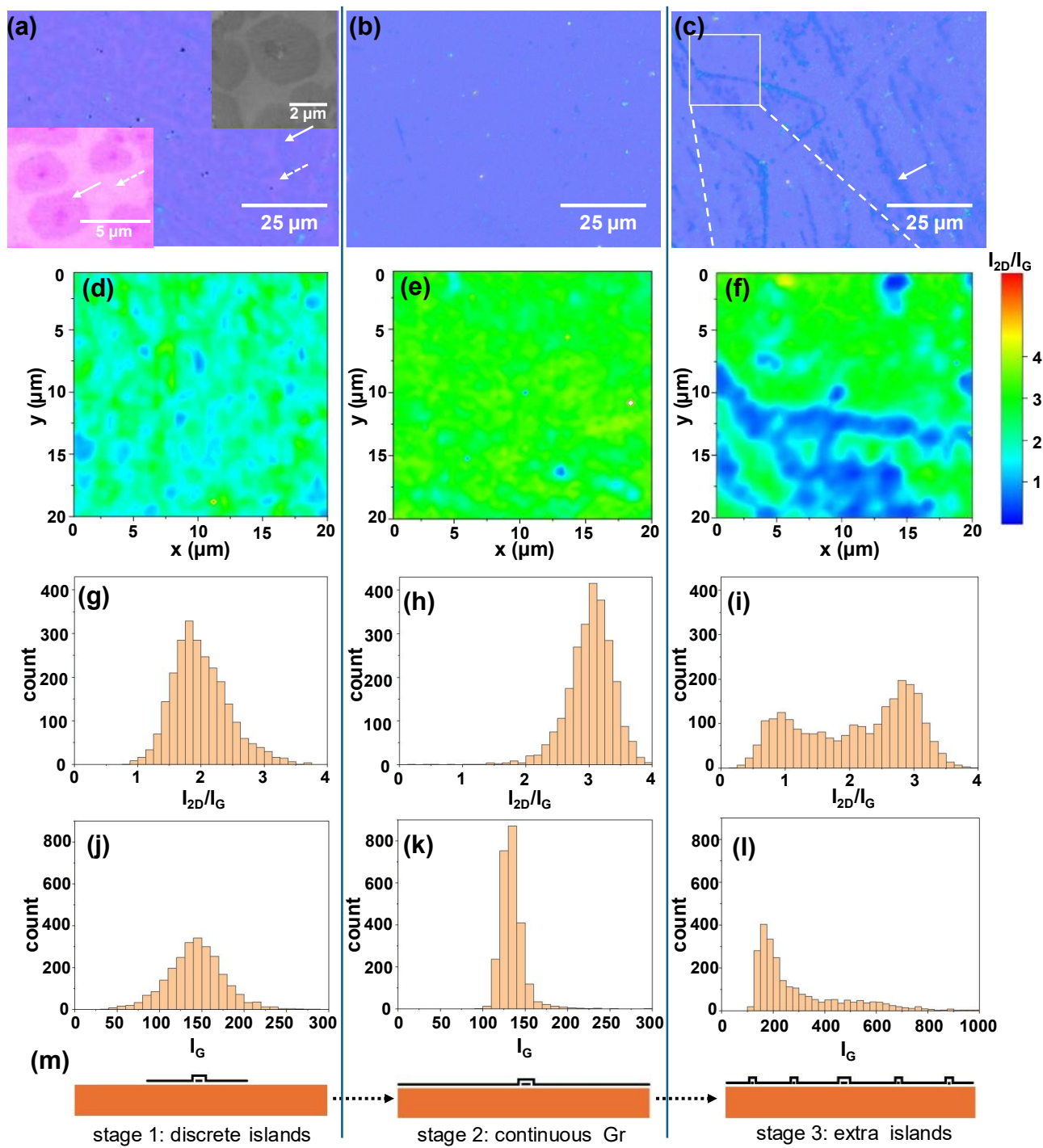

**Figure 2.** Characterization of Gr synthesized on a Cu foil using three different growth times ($t_{CVD}$): the 1st, 2nd, and 3rd columns are corresponding to $t_{CVD} = 10\ s$, $5\ min$, and $20\ min$, respectively. For detailed measurements, all Gr samples were transferred onto a silicon substrate with a 300-nm-thick silicon oxide layer (SiO₂/Si). (a)-(c) The optical

images of Gr on SiO$_2$/Si. The insets in the lower left and top right corner of (a) show the optical image with larger magnification and a low-voltage SEM image of the graphene islands, respectively. The bare SiO$_2$/Si substrate (purple color) and Gr covered region (blue color) are indicated by dashed and solid white arrows, respectively. (d)-(f) are the $20\ \mu m \times 20\ \mu m$ Raman mappings of the $I_{2D}/I_G$ ratio from the same samples in (a), (b), and (c), respectively. (f) is directly from the white square in (c). Based on (d)-(f), the statistics of $I_{2D}/I_G$ and $I_G$ are shown in (g)-(i) and (j)-(l). (m) The Gr growth mechanism via three different steps with increasing time: (left to right) discrete islands, a continuous layer, and a continuous layer with inserted islands at new nucleation sites, also known as the Stranski-Krastanov (SK) growth model [29].

As mentioned earlier, the Raman 2D mapping has limited spatial resolution and, therefore, the location-dependent characterization of Gr is still challenging. To address this, $I_G$ in **Figure 2**j-l can be used as it scales more sensitively with the number of Gr layers than $I_{2D}/I_G$ does [30]. Note that the $I_G$ distribution for $t_{CVD} = 5\ min$ is tighter (within $I_G = 100 \sim 180$) compared to the other two cases, again suggesting the weighted contributions of a uniform monolayer Gr is significant. In contrast, foil-based CGC with $t_{CVD} = 10\ s$ exhibits a broader $I_G$ distribution, ranging from about $50 \sim 250$. Lower values ($50 < I_G < 100$) can be linked to uncovered surfaces while larger values ($250 > I_G > 180$) can be explained by nucleation centers. Furthermore, $t_{CVD} = 20\ min$ results in $I_G > 100$ with a long tail up to about 1000, indicating the full surface coverage and strong location-dependent characteristics of Gr, e.g., the darker-blue strips in Figure **2**c and f (also see the high-$I_G$ portion in **Figure S6**c, supporting information). It appears that the locations of multilayer Gr overlap with the rolling patterns and grain boundaries in the Cu foil, known to be nucleation sites during CVD [25, 26].

In summary, our detailed analysis, using both optical/SEM images and Raman 2D mapping, concludes that the CVD Gr growth on Cu foil undergoes three distinctive steps, like Stranski-Krastanov (SK) growth. As schematically shown in **Figure 2**m (from left to right), the steps include (1) the nucleation of isolated Gr islands, (2) the formation of a continuous monolayer, and (3) insertion of additional Gr islands under the dominant monolayer Gr in the preferred locations for nucleation. The Cu surface acts as the catalyst for the decomposition of organic molecules, i.e., benzene, and the formation of Gr from free carbon adatoms. Because of the importance of Cu in CVD Gr growth, the growth of a second layer Gr on top of the continuous monolayer Gr becomes inefficient despite plentiful supply of benzene. As a result, additional nucleation sites have to be inserted between the dominant monolayer Gr and the Cu substrate (known as the inverse wedding cake growth mode) [18, 24, 31]. Note that the insertion as well as expansion of Gr under the dominant monolayer Gr are intrinsically slow because the supplies of additional carbon adatoms are mainly through defects in Gr after the full coverage of a Cu surface by continuous monolayer Gr [18].

**2.2 Effect of Gr continuity on electrical conductivity of CGCs**

In the previous section, we synthesized three different morphologies of Gr on Cu foils by tailoring CVD growth time. Here, the main focus is placed on investigating how these Gr morphologies are linked to the electrical conductivity of CGCs. For this, foil-, foam- and wire-based CGCs (see **Figure 1**) with different Gr morphologies were synthesized by precisely controlling the CVD growth conditions. For example, CVD growth time ($t_{CVD}$) was used as a main parameter for both foil- and foam-based CGCs while a benzene flow rate ($f_B$) was controlled for all wire-based CGCs (see Experimental Section for details). This modification for wire-based CGCs was important because finer wires became increasingly sensitive to prolonged thermal processes, e.g., excessive grain boundary sliding [12] (see **Figure S7** in the supporting information). To accurately quantify Gr-enhanced electrical performance, we prepared control Cu samples for all CGCs, i.e., pure Cu foils, wires, and foams underwent the identical thermal processes as their CGC counterparts but without being exposed to a benzene flow. In other words, the electrical conductivity of CGCs was directly compared to that of the corresponding annealed pure Cu matrix because thermal cycles under reducing atmosphere, e.g., during CVD, could also alter the electrical conductivity of Cu matrices. In addition, different types of Cu matrices from commercial vendors may have different intrinsic conductivity.

**Figure 3**a-c summarize the electrical conductivity of foil-, wire-, and foam-based CGCs compared to annealed Cu foils, wires, and foams, respectively (indicated as *Control* in the figure). For clarity, each CGC and its control in **Figure 3** are based on (a) 28-$\mu m$-thick foils, (b) 80-$\mu m$-diameter wires, and (c) 2-$mm$-thick foams. Detailed dimensions and electrical measurements for individual samples are given in **Table S1**-**3** in the supporting information. **Figure 3**d-f show the corresponding Raman spectra of CGCs to establish an experimental correlation between their conductivities and Gr morphologies where each curve is an average of multiple Raman measurements (see **Figure S8-S10** for more details). Furthermore, the electrical conductivity ($\sigma_s$) in **Figure 3**c indicates the conductivity of the constituents (or individual Gr-coated Cu ligaments) in a foam. It is well studied that $\sigma_s$ for the solid material comprising a high porosity metal foam [32-34] can be determined by:

$$\sigma_s = 3\rho_s \frac{\sigma_f}{\rho_f} \qquad (1)$$

where $\sigma_f$ and $\rho_f$ are the nominal electrical conductivity and the nominal density of a foam-based CGC, and $\rho_s$ is the density of a solid material comprising the high porosity foam, i.e., the Gr coated Cu material in the foam. In other words, Eq. 1 predicts the electrical conductivity of a highly porous metallic foam based on the material properties of the solid constituents. More details on Eq. 1 and the measured specific electrical conductivity ($\sigma_{Spec} = \sigma_f/\rho_f$) of foam-based CGCs can be found in Materials and Methods section and **Table S2** in the supporting information, respectively. For validation of Eq. 1, $\sigma_s = 58.3$ MS/m ($\approx 100.3\%$ IACS) was obtained for annealed Cu foams using the well-known bulk density of pure copper (8.96 $g/cm^3$) [7] and measured $\sigma_{Spec}$ for annealed Cu foams. It is worth noting that each copper matrix has slightly different intrinsic electrical conductivity (58.5, 57.0, and 58.3 $MS/m$ for Cu foils, wires, and foams, respectively).

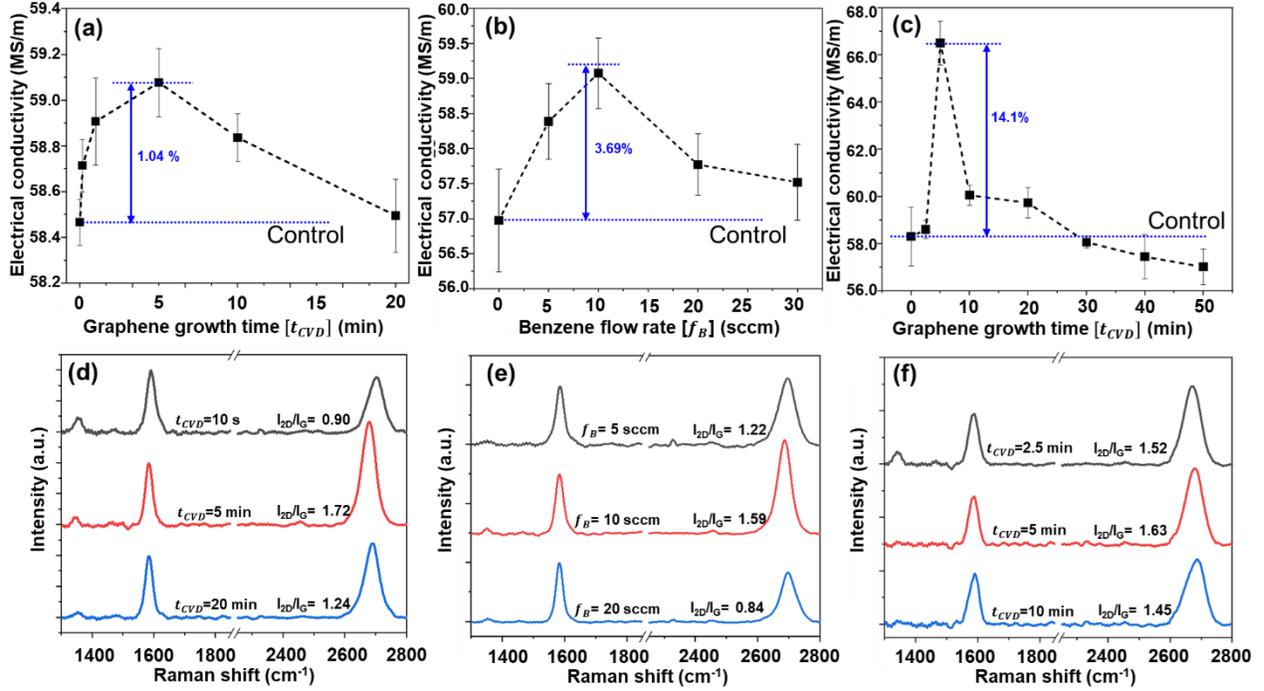

**Figure 3.** Electrical conductivities (a-c) and Raman spectra (d-f) of foil- wire- and foam-based CGCs (1st, 2nd, and 3rd columns, respectively). The electrical conductivities are shown as a function of CVD parameter (Gr growth time ($t_{CVD}$) in (a) and (c), and a benzene flow rate ($r_B$) in (b). In (a-c), *Control* indicates annealed Cu foils (about 28-$\mu m$ thick), wires (about 80-$\mu m$ in diameter), and foams (about 2-$mm$ thick).

For the electrical conductivity in **Figure 3**a-c, all three types of CGCs exhibit one similar trend — an initial increase, a peak value, and then a decrease. But each type achieves a significantly different maximum conductivity compared to each control (i.e., 1.04, 3.69, 14.1% improved conductivity for foil, wire, and foam-based CGCs, respectively). Three different CVD conditions for each type of CGCs were selected to capture an initial increase, a peak value, and then a decrease for further Raman analysis. For example, **Figure 3**d shows Raman spectra of foil-based CGCs at $t_{CVD} = 10\ s$, $5\ min$, and $10\ min$ because the highest conductivity occurs at $t_{CVD} = 5\ min$ and then the conductivity starts to decrease, e.g., $58.5\ MS/m$ at $t_{CVD} = 20\ min$. The average $I_{2D}/I_G$ ratios for $t_{CVD} = 10\ s$, $5\ min$, and $10\ min$ are 0.90, 1.72 and 1.24, respectively, indicating that the highest conductivity and largest $I_{2D}/I_G$ ratio correspond to each other. Another interesting observation is that $t_{CVD} = 10\ s$ results in a significantly higher $I_D/I_G$ ratio (= 0.25) compared to both cases for $t_{CVD} = 5\ min$ and $t_{CVD} = 20\ min$ ($I_{2D}/I_G < 0.1$), which can be attributed to pronounced Gr edges [27] from the isolated Gr islands (see **Figure 2**m, left). The conductivities of wire-based CGCs are 58.4, 59.1, and 57.8 $MS/m$ at $f_B = 5$, 10, and 20 $sccm$ (**Figure 3**b) with the average $I_{2D}/I_G$ ratio of 1.22, 1.59 and 0.84 (**Figure 3**e), respectively. Again, the highest conductivity and $I_{2D}/I_G$ ratio occur under the same CVD condition (i.e., $f_B = 10\ sccm$). Unlike foil-based CGCs, all wire-based CGCs exhibit a small $I_D/I_G$ ratio (< 0.1). Finally, the conductivities (**Figure 3**c) and the corresponding $I_{2D}/I_G$ ratios (**Figure 3**f) of foam-based CGCs are 58.6, 66.5, and 60.1 $MS/m$ and 1.52, 1.63 and 1.45 at $t_{CVD} = 2.5\ min$, $5\ min$ and $10\ min$, respectively. Again, this result concludes that the highest conductivity and $I_{2D}/I_G$ ratio are strongly correlated. The average $I_D/I_G$ ratios of foam-based

CGCs is 0.25 at $t_{CVD} = 2.5\ min$ but becomes much smaller ($< 0.1$) at $t_{CVD} = 5$ and $10\ min$ where the larger $I_D/I_G$ ratio can be attributed to edges of non-continuous Gr islands.

In summary, **Figure 3** indicates the strong correlation between the electrical conductivity and Raman spectra of CGCs for all three types of Cu matrices. This key experimental observation can be explained by the kinetics of CVD graphene growth – discrete Gr islands, continuous monolayer graphene, and insertion of discrete Gr (**Figure 2**m). The initial increase of electrical conductivity is likely associated with the transition from discrete Gr islands (**Figure 2**a) to continuous monolayer graphene (**Figure 2**b) because discontinuous features in Gr cause additional election scattering in CGCs [9]. All CGCs achieve the best electrical conductivity when $I_{2D}/I_G$ ratio is the highest, directly underscoring the importance of continuous Gr networks in CGCs to their electrical performance. The decrease in electrical conductivity is caused by the excessive insertion of discrete Gr islands between a Cu substrate and a continuous Gr layer (**Figure 2**c). Before the completion of the continuous layer, bare Cu surface without Gr will interrupt ballistic electron transport formed by Gr. After the completion of continuous Gr layer, insertion of discrete Gr deteriorates the electrical conductivity because (1) the small islands have lower electrical conductivity than the continuous monolayer [35]; (2) the adlayers actually introduces additional van der Waals gaps and interfacial resistance between the first layer and Cu, and therefore obstructs the electron transfer from Cu to the continuous first layer [36].

**Chemical robustness:** Here, we evaluate the anti-oxidation of CGCs after being exposed in an ambient condition for 4~6 months because chemical robustness is one of the desired properties for an effective conductor. Note that characterization of wire-based CGCs becomes non-trivial due to their small diameters and, therefore, foam-based and foil-based CGCs have been characterized by X-ray photoelectron spectroscopy (XPS) [37] and conductive atomic force microscopy (cAFM), respectively. **Figure 4**a shows the XPS intensity of foam-based CGCs using $t_{CVD} = 0, 2.5, 5,$ and $10\ min$. As shown earlier, the four CVD times represent control, discrete islands, a continuous layer, and a continuous layer with inserted islands. The control (or annealed Cu foam) exhibits a relatively higher O1s-to-Cu2p ratio, indicating poor oxidation resistance compared to all the other foam-based CGCs. **Figure 4**b shows the higher-resolution scan of Cu2p peak, where the characteristic satellites of $Cu^{2+}$ [37] was found only in annealed Cu foam. This result confirms that Gr in the CGCs serves as a protective diffusion barrier [38, 39] and, thus, achieves the effective anti-oxidation. Furthermore, the spectra of all the foam-based CGCs are smooth between the doublet Cu2p peaks of the CGC foams and their O 1s peaks become much smaller than the control. One interesting observation is that the O 1s peak for $t_{CVD}$= 10 min is larger than that of $t_{CVD}$= 2.5 and 5 min, which indicates that the former has more pronounced $Cu_2O$ level [40] than the latter. This conclusion implies that excessive insertion of discrete Gr results in lowering the effectiveness of the diffusion barrier compared to continuous monolayer Gr. **Figure 4**c-d shows the cAFM results of surface topography and the corresponding current map from a foil-based CGC with $t_{CVD} = 10\ s$. As a reminder, this sample corresponds to discrete islands (or a Cu foil partially covered by Gr). The voltage bias was applied to the foil in the out-of-plane direction (see **Figure S11**a in the supporting information for the schematic of cAFM). Isolated Gr islands can be clearly observed in **Figure 4**c. The current map in **Figure 4**d, by directly comparing with the patterns in **Figure 4**c, reveals that the regions covered by Gr has much higher current than the neighboring regions. This conclusion can be easily confirmed from the 3D maps in **Figure S11**b-c. For more quantitative analysis, **Figure 4**e and f compare the height and current profiles along three different paths (see the dashed lines). The vertical height of the Gr covered regions are about 6 nm lower than that of the uncovered regions in the topography map. Note that the Gr covered regions have at least $10^2$ higher current intensity than that of the uncovered regions. Interestingly, Gr ripples can be seen within a Gr island along Line 3. Despite considerable height variation (up to $5\ nm$ or about 40% change), the current value remains stable. During oxidation, Cu undergoes the volume expansion induced by lower density [7] and the adding mass of oxygen for oxide layer. Additionally, Cu oxide is nearly non-conductive under room temperature [7]. We therefore attribute the significant height and current difference between Gr coated and bare Cu surface to a Gr diffusion barrier. To prove this hypothesis, a foiled-based CGC with $t_{CVD} = 5\ min$ (i.e., continuous Gr) was also characterized by cAFM (see **Figure S11**d-f in supporting information). The fully covered CGC foil shows relatively homogeneous intensity in both topography and current map, no patterns of "basin" and "plateau" shown in the partially Gr covered Cu are observed. Additionally, by creating a pit after removing the surface material of $\approx 20\ nm$, the current intensity of the pit area, freshly created by an AFM tip, is in the same order as that of the Gr covered area, providing compelling evidence of the oxidation of bare Cu without Gr protection. These results suggest excellent anti-oxidation function of Gr coating on Cu for a relatively long time (6 months).

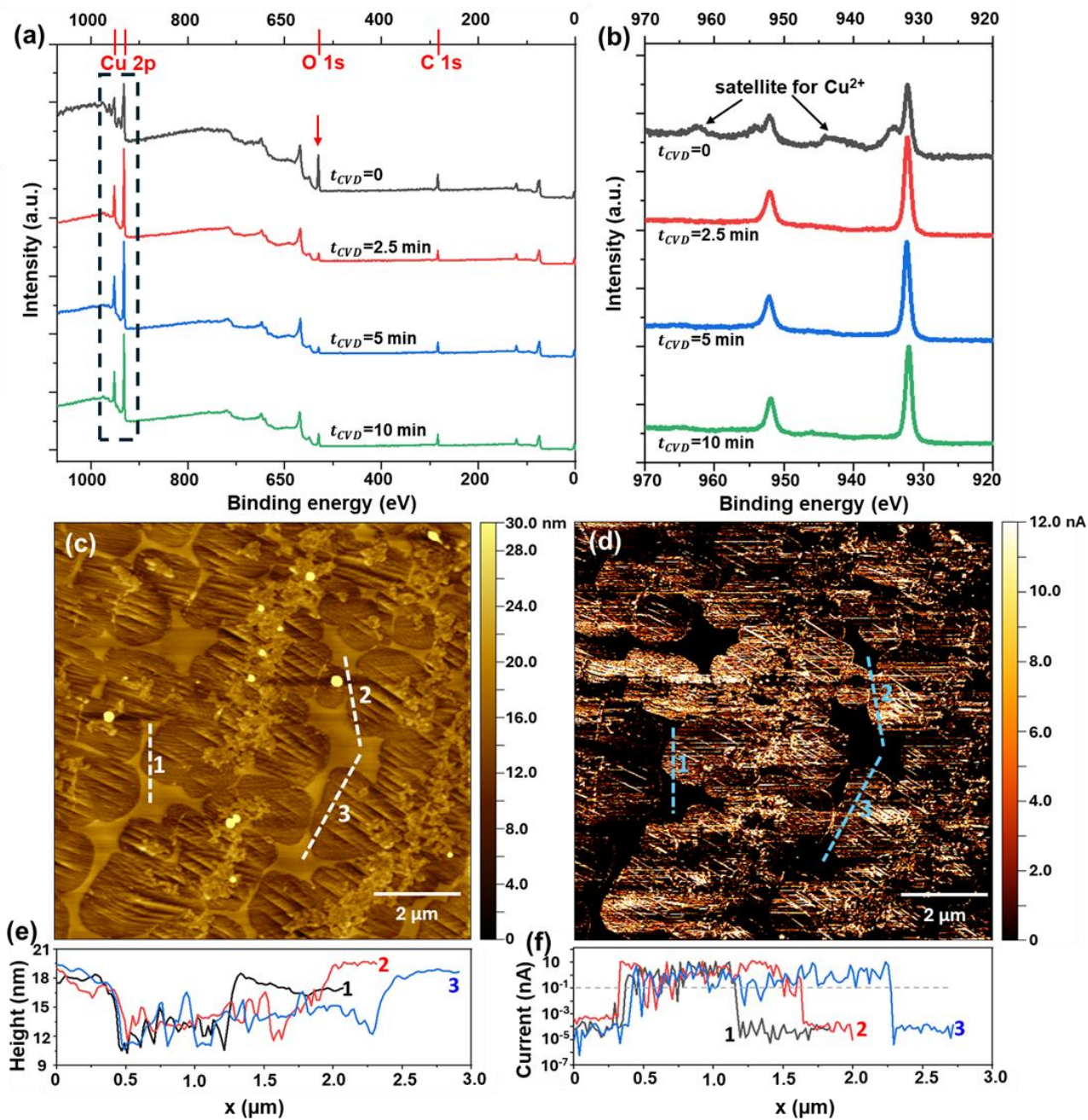

**Figure 4.** Anti-oxidation of CGCs. (a) and (b) characterization of surface composition for an annealed Cu foam (control) and the foam-based CGCs via X-ray Photoelectron spectroscopy (XPS). (a) The wide and (b) Cu2p (indicated in a dashed rectangle in (a)) scans of control ($t_{CVD} = 0$) and foam-based CGCs with $t_{CVD} = 2.5, 5,$ and $10\ min$, after storing all samples in an ambient air condition for 4 months. O1s peaks are indicated by a red arrow. (c) and (d) conductive atomic force microscopy (cAFM) results of a foil-based CGC with $t_{CVD} = 10\ s$ (isolated Gr islands), after storing samples in an ambient air condition for 6 months. (c) The morphology map ($10 \times 10\ \mu m^2$) of a randomly selected area and (d) the corresponding current map at $8\ mV$ bias for the same region. (e) and (f) the height and current profiles along three different paths indicated by dashed lines correspondingly in (c) and (d). The gray dashed line indicated in (f) shows the minimum detecting limit of the current measurement (current below this limit is often treated as zero nA).

### 2.3 The geometrical features of CGCs

In the previous section, we investigated the Gr-enhanced electrical conductivity in conjunction with kinetics of CVD graphene growth. The qualitative trend – the initial increase, peak value, and decrease of CGC conductivity – was explained by the three distinctive characteristics of graphene during CVD growth. However, this mechanism fails to explain the significantly different percentage changes achieved by each type of CGC (i.e., the largest conductivity improvements of foil, wire, and foam-based CGCs are 1.04, 3.69, 14.1%, respectively, compared to their control).

Here, we consider two key geometrical features (namely, a surface-to-volume ratio and cross-sectional shape) of a Cu matrix to explore the effect of these features on the maximum conductivity of CGCs. First, the surface-to-volume ratio, often quantified by the specific surface area ($A_s$ = surface area over mass), becomes important because Cu acts as catalyst during CVD Gr growth. In other words, it is reasonable to assume that Gr enhancement becomes more pronounced in CGCs with a larger Gr-to-Cu volume fraction. Second, the cross-sectional shape may determine the localized Cu-Gr interface and, therefore, influence how electrons locally interplay with the interface. As an example, the Cu-Gr interface with a large curvature may confine electron motion and likely reduce electron scattering within Cu.

**The correlation between the specific surface area and electrical conductivity of CGC:** We consider the change in electrical conductivity ($\Delta\sigma$) associated with the volume fraction of Cu and Gr ($X_{Cu}$ and $X_{Gr}$) in CGC by using the rule of mixture:

$$\Delta\sigma = \sigma_{CGC} - \sigma_{Cu} = (X_{Cu}\sigma_{Cu} + X_{Gr}\sigma_{Gr}) - \sigma_{Cu} \qquad (2)$$

where $\sigma_{CGC}, \sigma_{Cu}$ and $\sigma_{Gr}$ are the electrical conductivity of CGC, Cu, and Gr, respectively. Note that the volume of Cu ($V_{Cu}$) is significantly larger than that of Gr ($V_{Gr}$) because of atomically thin Gr in CGCs. Due to $V_{Cu} \gg V_{Gr}$, the volume fractions can be written as $X_{Cu} = V_{Cu}/(V_{Cu} + V_{Gr}) \approx 1$ and $X_{Gr} \approx V_{Gr}/V_{Cu}$ and, as a result, Eq. (2) can be simplified as

$$\Delta\sigma \approx \frac{V_{Gr}}{V_{Cu}}\sigma_{Gr} = \frac{A_{Gr}t_{Gr}}{V_{Cu}}\sigma_{Gr} = \rho_{Cu}A_s(\sigma_{Gr}t_{Gr}) \qquad (3)$$

where $\rho_{Cu}$ and $A_s$ are the density (8.96 $gcm^{-3}$) and specific surface area of the Cu matrix; $A_{Gr}$ and $t_{Gr}$ are the surface area and thickness of Gr. Eq. (3) implies $\Delta\sigma$ is linearly proportional to the specific surface area ($A_S$) of a Cu matrix and inversely proportional to the sheet resistance ($R_s = (\sigma_{Gr}t_{Gr})^{-1}$) of Gr.

For the complete analysis of the $A_S - \Delta\sigma$ relation observed in CGCs, we additionally prepared and characterized wire-based CGCs with both 10- and 25-μm diameters. It is important to note that these finer samples were fabricated by the identical CVD process of 80-μm-diameter wire-based CGCs with the highest conductivity (see **Figure 3**b). **Figure 5**a shows the average Raman spectrum of wire-based CGCs with three different diameters. The Raman signals of each CGC were collected from 10 different locations on wire surface as shown in **Figure S12**, the supporting information. The nearly negligible *D* peaks for all three different diameters indicate the low defect density and high coverage of Gr. One interesting observation is that the $I_{2D}/I_G$ ratio of wire-based CGCs appears diameter-dependent, despite use of the identical CVD parameters, as the ratio decreases from 1.59 to 1.02 with the reduction of wire diameter from 80 to 10 $\mu m$. This, while beyond the scope of the current work, can be explained in two possible ways. First, it is well known that Raman spectrum of Gr on a flat substrate is strongly influenced by laser incident angle [41, 42]. Note that surface curvature of fine wires can effectively change a local laser incident angle with respect to wire surface, especially when the laser spot size (i.e., $\approx 2.3\ \mu m$) is comparable to the wire radius (e.g., $r \approx 4.4\ \mu m$ for 10-μm-diameter wire). Second, the wires were produced by mechanical extrusions and, therefore, finer wires likely underwent more severe mechanical deformation. Such extreme mechanical deformation may introduce additional surface roughness, which can serve as additional nucleation sites.

The electrical conductivity of annealed Cu wires (control) and wire-based CGCs with three different diameters are summarized in **Figure 5**b. Detailed diameter and conductivity measurements for individual wire samples are given in **Table S4** in the supporting information. The electrical conductivity of wire-based CGCs with $d = 80$, 25, and 10 μm are 59.07, 59.41 and 59.32 $MS/m$, respectively, all higher than IACS (58.1 $MS/m$). The improvement in the apparent electrical conductivity ($\Delta\sigma$) of wire-based CGCs compared with their control are 3.69% (2.10 $MS/m$), 12.2% (6.45 $MS/m$) and 17.1% (8.67 $MS/m$), respectively. Obviously, $\Delta\sigma$ increases with the reduction of wire diameter, confirming the effect of the specific area on the overall electrical conductivity of CGCs.

The specific surface area ($A_S$) of the foil-, wire- and foam-based CGCs was experimentally determined to estimate their volume fraction of Gr ($X_{Gr}$). For foil and wire cases, their simple geometries allow us to measure their thickness and diameters to calculate their $A_S$. For wire, the measured diameters after CVD were 8.73, 22.48, and 77.33 $\mu m$ for nominally 10-, 25-, and 80-$\mu m$ diameter wires and, as a result, $A_S = $ 0.0512 and 0.0199 and 0.00577 m²/g, respectively. Similarly, the thickness and specific surface area of foil-based CGCs were 27.81 $\mu m$ and 0.00803 $m^2/g$. To capture the complex geometry of foam-based CGCs, 3D micro-CT scan was conducted. **Figure 5**c1 and c2 show the overview of a foam and the zoom-in view of a ligament, respectively. **Figure 5**c3 shows the cross sections of the ligament in **Figure 5**c2 at four different locations from ① to ④ (see **Figure S13** in the supporting information for additional cross-sections) where ① and ④ form a junction with neighboring ligaments. The cross-sections near the ends of the ligament (e.g., ① and ④) tend to have a larger hole and a thinner wall thickness than that of the middle part (e.g., ② and ③) (also see **Figure 1**d). Image analysis indicates that the perimeter-to-area ratio is in the range of $0.1790$~$0.2452\ \mu m^{-1}$ or equivalently the specific surface area of $0.0200$~$0.0274\ m^2/g$. It is worth noting that a Langmuir isotherm method [43] measured $A_S = 0.0243\ m^2/g$ for bulky CGC foam samples (see **Figure S14** and **Table S5** in the supporting information for details).

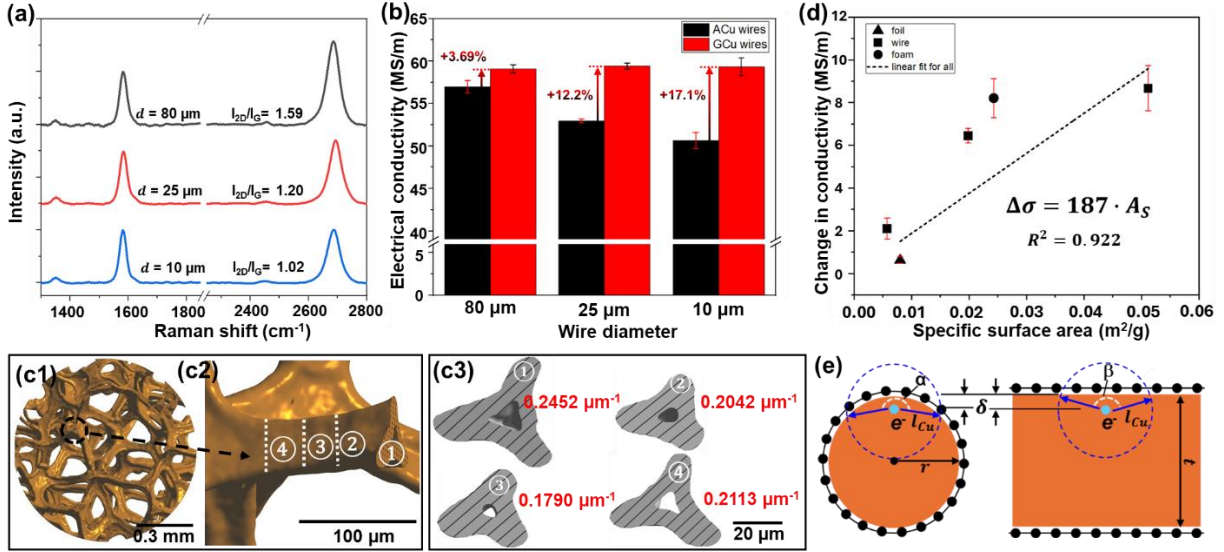

**Figure 5.** Gr-enhanced electrical conductivity in three different types of CGCs. (a) The Raman spectra and (b) electrical conductivity of wire-based CGCs and their controls (annealed Cu wires) with nominal diameters of 80 μm, 25 μm and 10 μm. (c) the 3D image of a foam-based CGC using micro-computed tomography. (c1) an overview of a 3D foam, (c2) a zoom-in view of a ligament (circled in (c1)) and (c3) cross-sections of the ligament from ① to ④ indicated in (c2). The value in red color is the perimeter to the area ratio of each cross section. (d) The improvement of electrical conductivity ($\Delta\sigma$) of foil-, wire-, and foam-based CGCs as a function of their specific surface area ($A_s$). Note that $\Delta\sigma$ is with respect to the annealed pure Cu counterpart of each CGC. The black dashed line is the result of a linear fit for the $\Delta\sigma$-$A_s$ relation with $R^2 = 0.922$. (e) Electron confinement for an electron ($e^-$) located at $\delta$ from the Cu-Gr interface: (left) a wire-based CGC and (right) a foil-based CGC. $r$ is a radius of a wire and $t$ is the thickness of a foil. The dashed circles represent a radius of the mean free path ($l_{Cu}$) of an electron within Cu matrices. $\alpha$ and $\beta$ are the critical angle within which an electron travels back to the Cu-Gr interface. Note that $\alpha > \beta$ due to the different geometries of the two GCGs.

**Figure 5**d summarizes the experimentally measured $\Delta\sigma$–$A_s$ relation of all CGCs where $\Delta\sigma = 187 A_s$ with $R^2 = 0.922$ (see a dotted line) is obtained based on a linear fitting with y-intercept set to be 0. This result implies that Eq. 3 captures a significant portion of the experimentally measured $\Delta\sigma$–$A_s$ relation. Furthermore, the 10-$\mu m$-diameter wire-based CGCs achieve the largest Gr-enhanced electrical conductivity (17.1% improvement) due to the highest $A_s$ value. Note that $\Delta\sigma$ values for the 80- and 25-μm-diameter wire-based CGCs are located above the dashed line unlike the 10-$\mu m$-diameter wire-based CGCs. This indicates that the sheet resistance of Gr in 10-$\mu m$-diameter wire-based CGCs is the highest among the wire-based CGCs. This can be explained by different Gr qualities as evidenced by $I_{2D}/I_G$ values in **Figure 5**a although all wire-based CGCs were synthesized under the identical CVD conditions. This seemingly size-dependent Gr growth on a Cu wire remains as a future work. Another key conclusion is that bulk-scale foam-based CGCs also follow Eq. 3, highlighting the promising potential for manufacturing effective bulk-scale Cu-Gr conductors.

Using Eq. 3, $\rho_{Cu}/R_s = 187\ MS \cdot gm^{-3}$ can be calculated, resulting in $R_s = 0.048\ \Omega/sq$. Assuming continuous monolayer Gr in CGCs, the effective conductivity of Gr at the Cu-Gr interface becomes $\sigma_{Cu-Gr}^{Eff} = (t_{Gr} R_s)^{-1} = 6.23 \times 10^4\ MS/m$ where Gr thickness is $0.335\ nm$. Note that $\sigma_{Cu-Gr}^{Eff}$ is significantly higher than the intrinsic conductivity of the high-quality pristine Gr ($\approx 100\ MS/m$) as well as IACS (about 1000 times higher) likely due to electron doping of Gr from Cu [11]. It can be argued that Cu can donate electrons to Gr without altering the electronic properties of a Cu matrix because Cu has an extreme high carrier density ($8.5 \times 10^{28}\ m^{-3}$) [7], four orders of magnitude higher than that of Gr ($6 \times 10^{24}\ m^{-3}$) [6], and an extreme small Cu-to-Gr volume fraction.

**The effect of the cross-sectional shape on electrical conductivity of CGC: Figure 5**d shows that foil-based CGCs and 80-μm-diameter wire-based CGCs exhibit different enhancement ($1.04\%$ versus $3.69\%$) although the former has higher $A_s$ ($0.00803\ m^2/g$) than the latter ($0.00577\ m^2/g$). This result may indicate that the conductivity of CGCs is sensitive to the shape of their cross sections. To explain this, consider a cross section of wire- and foil-based CSCs in **Figure 5**e. In each CGC, an electron is located at a small distance $\delta$ away from the Cu-Gr interface where the dotted circle indicates the mean free path of an electron in Cu, designated by $l_{Cu}$. Note that $\alpha$ and $\beta$ are the angle between the intersections of the Cu-Gr interface and mean free path with respect to the electron. From the given geometry, it is obvious to show $\alpha > 180° > \beta$ and, as a result, a randomly scattered electron in the wire-based CGC has higher probability to move to the Cu-Gr interface compared to that in foil-based CGC. This simple argument suggests that electrons can be more effectively accumulated at the Cu-Gr interface when an electron is geometrically confined by the large curvature, e.g., observed in wire- and foam-based CGCs.

## 3. Conclusion

This work investigates the underlying mechanisms of the Gr-enhanced electrical conductivity in copper-graphene composites (CGCs). For this, we synthesized and characterized the different types of Gr-coated Cu composites to control the characteristics of both Gr and Cu in CGC. For example, we performed careful parametric studies on CVD Gr growth on three types of Cu substrates, i.e., foils, wires and foams. By utilizing this innovative experimental approach, we unveiled the key mechanisms of the Gr-enhanced electrical conductivity of CGCs: the Gr-enhancement strongly depends on the continuity and quality of graphene as well as the geometry of a Cu matrix.

We have identified three different configurations of CVD-grown Gr on Cu substates, namely, isolated islands, continuous monolayer, and insertion of islands. Gr growth starts from the formation of small, isolated islands. Note that our experimental analysis has shown that the center of each island is typically bi- or multilayer Gr while the rest is covered by a monolayer. Then the dominant monolayer laterally expands until a continuous Gr forms. With excessive CVD growth, additional nucleation sites form and the size of the embedded Gr layer under the dominant monolayer slowly increases because Gr growth is preferable on a Cu catalyst.

The capability to control the characteristics of Gr allows us to quantitatively characterize the effect of the Gr continuity and quality on CGC performance. Our conductivity measurements clearly indicate that continuous monolayer graphene in CGC results in the best electrical performance, e.g., an unprecedented 17.1% improvement compared to pure Cu counterpart, among the three configurations of Gr. This result can be explained by the ballistic electron transport along continuous high-quality graphene. On contrary, isolated Gr islands likely interrupt effective electron pathways due to insufficient Gr coverage while excessive insertion of Gr islands increases the interface resistance.

Furthermore, we have utilized three different types of Cu substrates to investigate the effects of Cu geometries, such as surface-to-volume ratios and cross-sectional shapes, on the electrical conductivity of CGCs. Utilizing foil-, foam-, and wire-based CGCs, 1.04%, 14.1%, and 17.1% higher conductivities compared to their Cu counterparts have been unambiguously achieved. Two mechanisms for this unexpected result have been explored. First, the electrical conductivity improvement ($\Delta\sigma$) strongly depends on the specific surface area ($A_s$) of a Cu matrix because a larger volume fraction of Gr in CGC increases the contribution of Gr, giving rise to the strong $\Delta\sigma \propto A_s$ relation. Also, Gr enhancement is more pronounced when an underlying Cu matrix has a curved cross-section (e.g., a wire or ligaments of a foam compared to a foil) likely because electron motions are more tightly confined by Gr.

In summary, this study has unambiguously demonstrated that Gr can significantly enhance the electrical conductivity of CGCs only when the characteristics of both Gr and Cu are optimally tailored. Our results provide a reasonable explanation for a broad range of the electrical conductivities of CGCs reported in the literature [9, 10]. Also, this work can serve as a reference for guiding future material design and mass production of effective CGC conductors. Some examples of future CGC applications may include CGCs with specifically tuned electrical conductivity and roll-to-roll manufacturing of high-performance CGC conductors using macroscopic Cu-foam based CGC.

## 4. Materials and Methods

***Fabrication of the Graphene Coated Foams:*** The as-received Cu foams with thickness of 2 mm (Futiantian Technology Co.) were cut into pieces of 4 cm by 7.5 cm and then loaded into the home-made CVD system (**Figure S15**a, supporting information) as the substrates for graphene growth. The 3-step CVD process (**Figure S15**b, supporting information) included ramping, soaking and cooling, in which the temperature, gas flow rate and CVD time could be adjusted based on the properties of different substrates. The Cu foams were first ramped from room temperature (20 °C) to 1000 °C in 40 $min$ with a uniform heating speed under a gas flow of 100 standard cubic centimeter per minute ($sccm$) of Ar and 10 $sccm$ of H$_2$. The ramping was followed by an 80 $min$ soaking under 1000 °C with a gas flow of 1500 $sccm$ of Ar and 100 $sccm$ of H$_2$. The benzene vapor (Aldrich, 99.8%, anhydrous, 55 °C heating by a heating plate) with a flow rate of 5 $sccm$ was introduced into the CVD chamber as the gaseous carbon source during the last $x\ min$ ($x = 2.5, 5, 10, 20, 30, 40, 50$) of soaking. The foam-based CGCs with different CVD growth time ($t_{CVD} = x$) were coated by graphene with various quality and different coverage. The annealed Cu foams were prepared with the same thermal history and gas flow with the Gr coated samples without introducing benzene (i.e., $t_{CVD} = 0$).

***Fabrication of the Graphene Coated Foils and Wires:*** Cu foils (99.99%, Chudeng Ltd.) of a nominal thickness of 30 μm were cut into strips of 5 $mm \times 100\ mm$ and ultrasonic cleaned by isopropanol and acetone successively. The Cu strips were then soaked into DI water for 3 times and dried in air before being wrapped (**Figure S15**c, supporting information) on the Cu sample frame. The samples on the frame were then coated with Gr by CVD with the same process as the foams, except lowering the benzene vapor flow rate ($f_B$) to 1 $sccm$. The Gr coated foil sample has $t_{CVD} = 10\ s, 1\ min, 5\ min, 10\ min, 20\ min$, respectively. The annealed Cu foils were prepared with the same thermal history and gas flow with the Gr coated foil samples without introducing benzene (i.e., $t_{CVD} = 0$). Cu wires with diameter 80 μm (99.99%, California Fine Wire Co.) were wrapped onto a nickel sample frame (**Figure S15**d, supporting information) and loaded into the CVD system. Before the 3-step CVD, the samples were pre-annealed under 750 °C with 1500 $sccm$ Ar and 100 $sccm$ H$_2$ for 40 minutes, followed by a soaking under 200 °C with 760 $sccm$ Ar and 30 $sccm$ H$_2$ for 30 minutes, to remove the pre-existing oxide and impurities on the surface of Cu wires. After the pre-annealing, the wires were ramped from room temperature (20 °C) to 960 °C in 30 min with a uniform heating speed under a gas flow of $760\ sccm$ of Ar and $30\ sccm$ of H$_2$. The soaking time under 960 °C was $25\ min$ and the benzene was introduced at the last $10\ min$ with $f_B = 5, 10, 20,$ and $30\ sccm,$ respectively. The annealed Cu wire counterparts were prepared with the same process without benzene gas flow. Gr coated and annealed wires with nominal diameter 25 and 10 μm (99.99%, California Fine Wire Co.) were prepared by the same process as that for 80 μm wires, with $f_B = 10\ sccm$.

***Handling and Characterization of Different Types of CGCs:*** The electrical resistance of the samples was measured by four probe tests (**Figure S16**a**,** b, supporting information) with Keithley DM6500 multimeter. Foam-, foil- and wire-based CGCs were cold mounted and polished for OM of their cross-sections. The Raman spectroscopy with laser wavelength 532 nm was employed to characterize the Gr.

*Foam-based CGC:* The CGC foams were gently cut into strips with a width of $5\ mm$. The thickness of the foams was measured by a modified digital micrometer (Mitutoyo micrometer, **Figure S16**c, supporting information). The Si wafer attached to the head of the micrometer enlarged the contact area of the foam and decreased the pressure applied to the form to avoid possible deformation of the foam. The length and width of the foams were measured by using a Mitutoyo digital caliper. The weight of the foams was measured by an OHAUS analytical balance with a precision of $0.0001\ g$. The density, porosity, electrical conductivity of the foams was then calculated based on the electrical resistance, dimensions, and mass. After 4 months of the synthesis of foam-based CGCs, the oxidation level of the foams was characterized by XPS (Kratos Axis Supra +). The specific surface area of the foam was measured by Tristar II Plus with N$_2$ being the adsorption gas.

*Foil-based CGC:* The thickness of each foil was determined based on their measured planar dimensions and mass. The Mitutoyo digital caliper and OM (KEYENCE Digital Microscope VHX X1) were used to measure the planar dimensions. The cross-section of CGC foils were carefully prepared (see **Figure S17**, supporting information), showing good thickness uniformity along the axial direction of the foils. The average foil thickness for each sample was then calculated by dividing the measured mass by the product of the Cu density and foil's planar area. Gr coated on the surface of the foils were transferred to $300\ nm$ SiO$_2$/Si [44, 45] for optical images and Raman mapping (see **Figure S3**, supporting information). 4.5% PMMA (average molecular weight 120000, Sigma Aldrich) was dissolved in chlorobenzene (99%, Sigma Aldrich) and spin coated on the surface of foil-based CGCs. The samples were then baked at 80 °C for $15\ min$. The Cu substrates were then etched away by $0.5\ M$ FeCl$_3$ solution. The PMMA-coated Gr samples were scooped out by SiO$_2$/Si and then baked under $80 - 130$ °C for $25\ min$. The PMMA layer was removed by hot acetone. After 6 months of the synthesis of foil-based CGCs, the surface conductivity of the foils was characterized by cAFM (Bruker Dimension Icon) in the ambient environment by NCH PtSi conductive probe having high spring constant (Nominal $k = 42\ N/m$). The cAFM image of **Figure 4** was captured at $8\ mV$ bias with a probe setpoint pressure of $0.13\ V$, $0.9\ Hz$ scan rate and 512 samples/line. For **Figure S11** in supporting information, at first, $1 \times 1\ \mu m^2$ area was scanned at 128 samples/line) by applying slightly high pressure (setpoint $0.5\ V$) to remove the material at a controlled removal rate of $5\ nm$ per scan. Once the $20\ nm$ depth removal was done, a comparatively larger area, $3 \times 3\ \mu m^2$ was scanned with gentle pressure (setpoint $0.12\ V$) while applying $3\ mV$ bias between the sample and AFM tip for current mapping. The scan rate was $0.9\ Hz$ and samples/line was 256.

*Wire-based CGC:* The geometry of the wires was observed under SEM (Helios 5 UX). The cross-section of the wire-based CGCs were carefully prepared and observed under KEYENCE Digital Microscope VHX X1 (see the insets in **Figure 1**b-d**)**, showing the samples maintain an essentially circular geometry. SEM images of various magnification confirm the structural uniformity of the CGC wires (see **Figure S18**, supporting information**)**. The average cross-sectional area was calculated by the circular equation based on the average diameter measured for each wire under SEM (Helios 5 UX). To improve the reliability of the diameter measurements, $50 - 100$ sections were selected along the entire CGC wire for each sample to ensure high enough sampling density.

*Micro-CT and 3D modeling:* The 3D microstructure of a $1\ mm \times 1\ mm \times 1\ mm$ cube from a CGC foam sample was obtained by micro-CT scanning with ZEISS Xradia 520 Versa. Micro-CT imaging combined with 3D Slicer software is used to reconstruct and analyze the 3D structure of copper foams. Contrast enhancement and noise reduction are implemented on raw micro-CT images to improve segmentation accuracy. The software applies filters to amplify edge detection and remove artifacts and ensure precise structural analysis. This segmentation method is performed under threshold-based or machine-learning-based process to visualize copper foam structure. Additionally, the thresholding procedure differentiates between pore space and solid copper by adjusting grayscale intensity values. The final foam's morphology is created with rendering segmented images, providing a comprehensive visualization of the 3D model. This model enables detailed analysis of overall structure integrity, distribution, and pore connectivity (see Supporting information for details).

**Electrical conductivity of foam:** To characterize the electrical performance of the annealed and CGC foams, the specific electrical conductivity (the electrical conductivity normalized by density) is measured because of the following reasons. The apparent electrical conductivity of the as-prepared foam is a joint consequence of the intrinsic electrical property of the constituent (the solid material) and the spatial distribution of the constituent (i.e. porosity). To understand the intrinsic electrical properties of the foam, the impact of the porosity in the foam must be ruled out. The specific electrical conductivity ($\sigma_{Spec}$, the electrical conductivity normalized by density) of the as-prepared foam is proportional to the specific electrical conductivity of the solid material comprising it by formulated as equation (4) [46]:

$$\sigma_{Spec} = \frac{\sigma_f}{\rho_f} = \frac{1}{3} \cdot \frac{\sigma_s}{\rho_s} \qquad (4)$$

Here $\sigma$ and $\rho$ are electrical conductivity and density where subscripts $f$ and $s$ denote the foam and the solid constituents comprising the foam, respectively. The second part of equation (1) links the apparent, measurable properties of the foam to the properties of the solid materials comprising the foam that are not readily measurable, and the ratio $\frac{1}{3}$ works specifically for high-porosity open-cell metal foam [32-34]. Justified by equation (4), the porosity-independent electrical property of Gr coated Cu material ($\sigma_s$) in CGC foam can be simply inferred from the specific conductivity ($\sigma_{Spec}$) of the CGC foam. The $\sigma_{Spec}$ of annealed Cu foams was measured as $2.17\ Sm^2/g$. By applying equation (4) and substituting $\rho_s = 8.96\ g/cm^3$, the $\sigma_s$ of solid annealed Cu was calculated to be $58.3\ MS/m$ ($\approx 100.3\%$ IACS), which was a decent match with the literature value. Nevertheless, we still chose Cu foam pieces with density within $\pm 1\%$ of average to minimize any possible impact from porosity (see **Figure S19** in supporting information).

**Supporting Information**
Supporting information is available from the Wiley Online Library or from the author.

**Acknowledgments**

This project was supported by the National Science Foundation (NSF CMMI-2338609, CAREER award) and the Office of Naval Research (N00014-23-1-2388, Dr. Paul Armistead).

**Conflict of Interest**

The authors declare no conflict of interest.

**Data Availability Statement**

The data that support the findings of this study are available from the corresponding author upon reasonable request.